\documentstyle[12pt,aaspp4,astrobib,epsfig,here]{article}

\def\Msun{\rm M_\odot}

\def\ltsima{$\; \buildrel < \over \sim \;$}
\def\simlt{\lower.5ex\hbox{\ltsima}} 
\def\gtsima{$\; \buildrel > \over \sim \;$}
\def\simgt{\lower.5ex\hbox{\gtsima}} 

\begin{document}

\title{Was GRB~980329 at $z \sim 5$?}

\author{Andrew S. Fruchter}
\affil{Space Telescope Science Institute, 3700 San Martin
Drive, Baltimore, MD 21218, USA}

\begin{abstract}
The optical transient (OT) associated with GRB~980329 was
remarkably red.  
It has previously been concluded that this was the result
of 
dust extinction in the host galaxy \cite{tfk+98,rlm+98,ppm+98}.  
However, an extinction model 
can only agree with the data if the $I$ band
observations taken about 0.8 days after outburst are discounted
\cite{kml+98,rlm+98}; the flux density ratio between the $I$  and
the $R$  of a factor $\sim 7$ is too great to 
be explained by extinction, given the relatively blue
$K-I$ color. 
Here it is shown that the {\it entire} observed optical/infrared spectrum is
consistent with that which is expected
from an unextincted OT at $z \sim 5$.  
At this redshift, the light in
the observer's $R$ band is
strongly suppressed by absorption 
in the Ly$\alpha$ forest -- an effect which 
has been clearly seen in galaxies 
in the Hubble Deep Field  \cite{wsb+98,ssb+98}. In
spite of its potentially high redshift, GRB~980329
was an unusually bright burst.
If GRB 980329 was indeed at $z \sim 5$, and its gamma-rays were
radiated isotropically,  the implied
energy of the burst would be $ 5 \times 10^{54}$~ergs.  
Should GRB~980329 have a host galaxy, deep imaging could 
confirm or reject the conclusion that this burst was at $z \sim 5$.
\end{abstract}

\section{Introduction}

The last eighteen months have seen a dramatic transformation in our
understanding of the nature of gamma-ray bursts (GRBs).  The discovery
of OTs associated with GRBs ({\it e.g.} \citeNP{vgg+97,bond6654}) 
has led in turn
to direct proof that these objects are at cosmological distances 
({\it e.g.} \citeNP{mdk+97,kdr+98}).  
Analyses of the spectra and temporal behavior of
the afterglows from the radio to the x-ray \cite{fkn+97,wrm97,gwb+98}
have confirmed that we are observing expanding relativistic fireballs
\cite{goo86,mr97}.   Nonetheless, the ultimate astrophysical source
of GRBs remains obscure.

The most widely discussed mechanisms for producing GRBs
-- binary neutron-star or black hole - neutron star mergers, and
the collapse of massive stars \cite{elps89,npp92,bp98} 
-- should be associated
with star formation, and indeed a number of well-studied
GRB hosts do show signs of intensive star-formation 
\cite{fpt+98,dkb+98,bdk+98}.   Additionally, the distribution of host
galaxy magnitudes is consistent with a model that links the
rate of GRBs at a given epoch of time with the star-formation
rate at the epoch \cite{hf98}.  But it is the location of
the GRBs in their host galaxies which one might expect
to distinguish between binary merger and massive stellar collapse,
also know as hypernovae, models \cite{bp98,lsp+98}.  
Binaries containing a neutron star should travel far from their birthplace
before merger due to the kicks neutron stars receive at birth
both from the loss of mass in the binary, and from possibly large
impulses from the supernovae \cite{dc87,ll94}.  Indeed, neutron star binaries
may frequently escape their host galaxy before merger \cite{bsp98},
preventing the creation of an afterglow and perhaps
even the GRB itself, due to the absence of a dense
external working surface \cite{mr93,sp97}.   However, GRBs caused
by hypernovae would be expected to occur at the locations
of star formation, and thus might be frequently enshrouded
in dust \cite{bp98}.    

Although the optical
colors of some OTs have shown evidence of moderate extinction
\cite{rei98,kdr+98,fpt+98}, and the inability of observers to find
OTs for some GRBs has been interpreted as possible evidence
of dust obscuration \cite{ggv+98,tfk+98,bp98}, 
there is little direct evidence that GRBs occur
in regions of high extinction.
GRB~980329, however,
appeared to be an excellent candidate for a dust enshrouded GRB.  Its OT was 
not discovered until a radio identification \cite{tfk+98}
allowed optical observers to re-examine images taken on
the first night after outburst.   The first image to show
an OT was an uncalibrated $I$-band image \cite{kml+98}.
The derived magnitude $I \sim 20$ at 0.8 days after outburst
 was surprising given that a
much deeper $R$-band image taken at the same time \cite{ppm+98}
found the source to have $R = 23.6 \pm 0.2$.  Although the $I$-band value
has now been calibrated and its brightness lowered to $I = 20.8 \pm 0.3$
(this value will appear in a forthcoming version of Reichart et al. 1998)
the $R-I$ color, 2.8,  is extremely red.
Subsequent imaging in $K$, $J$ and $R$ has shown that this OT also has 
very red $K-R$ and $J-R$ colors (see Reichart
et al. 1998 for a clear and comprehensive review of the observations of
this object; a complete table of the optical and near-infrared observations
on this object is also available in Palazzi et al. 1998).
While the very red $K-R$ and $K-J$ colors can be explained by strong extinction,
it is difficult to reproduce the very steep $R-I$ color by such a model,
given that the observed $K-I$ is quite blue. 
Indeed, in order to do so, one must either
discount the $I$ band image, or stretch the errors to their limits 
\cite{rlm+98}.  
Here, it is assumed that the data are correct and not misleading. 
To fit the data, a radical but plausible alternative is proposed: 
GRB~980329 occurred at a redshift of $\sim 5$.

\section{Analysis}

Although measurements of the OT of GRB~980329 were 
taken in the $K$, $J$, $R$ and $I$ bands,
on no single day are good data available in more than two of these bands
(see Table 1 of Reichart et al. for a complete list of the available 
observations). 
The standard method of dealing with this situation would be to determine
the index of the power-law decay of the OT, and use this to interpolate
to a common time.  However, as will shortly be shown, the crucial observation
was the $I$ band detection of the OT, and the object
was observed in $I$ only on the first night.
We are therefore forced to extrapolate
the later time $K$ and $J$ observations back to the first night.  To do this,
we use two methods, but both rely on the assumption
that
the flux density in the different bands falls with the same temporal
power law.
This assumption should be correct so
long as a spectral break in the synchrotron emission of the OT
did not pass through the waveband of interest  between the times of the
different observations used for the analysis.  As noted by Reichart et al. (1998)
the  power-law indices obtained for the $R$-band data, $-1.29 \pm 0.19$,
and the $K$-band data, $-0.98 \pm 0.30$, agree within the errors.
Since these two bands represent the short and long wavelengths
ends of the spectral region of interest, the assumption of a single
power-law index should be good for all the data.  Furthermore,
the average of these two values  $-1.14 \pm 0.15$
agrees well with the power-law indices determined for a number of
bursts \cite{fpt+98,pfb+98,hth+98}.

If one then employs the average power-law index of $-1.14$ to extrapolate
the $J$ and $K$ magnitudes back to the time of the $I$ and $R$ images
on the first night, 29.9 March 1998, one finds
estimated magnitudes of $K = 18.7 \pm 0.4$ and  $J = 20.4 \pm 0.4$.
As a check on these numbers, one can also subtract the observed 
$R-K$ color determined
at the one time
when both bands were observed,
2.3 April 1998, from the observed $R$ magnitude for 29.9 March 1998. One
then finds a $K$ magnitude of $18.5 \pm 0.4$, which is, within the errors,
identical to the $K$ magnitude estimated previously.  If one additionally
adds the $J-K$ color of 6.3 April 1998 to the $K$ magnitude derived
for 29.9 March 1998, one
again arrives at a $J$ magnitude of $20.4 \pm 0.4$.  For the
remainder of the paper, then, the values of $K = 18.6 \pm 0.4$ and
$J =20.4 \pm 0.4$ will be assumed for 29.9 March 1998.
These derived magnitudes have been corrected for 
Galactic foreground extinction, converted to $\mu$Jy and
plotted in Figure~1.   The $I$ band magnitude used is this figure
is the recalibrated value ($I = 20.8$) determined by Reichart et al.
The foreground Galactic
extinction  used, $E(B-v) = 0.074$, 
is that predicted by the  100~$\mu$ IRAS maps following
the method of Schlegel, Finkbeiner and Davis (1997). \nocite{sfd98} 

\begin{figure}
\centerline{\psfig{file=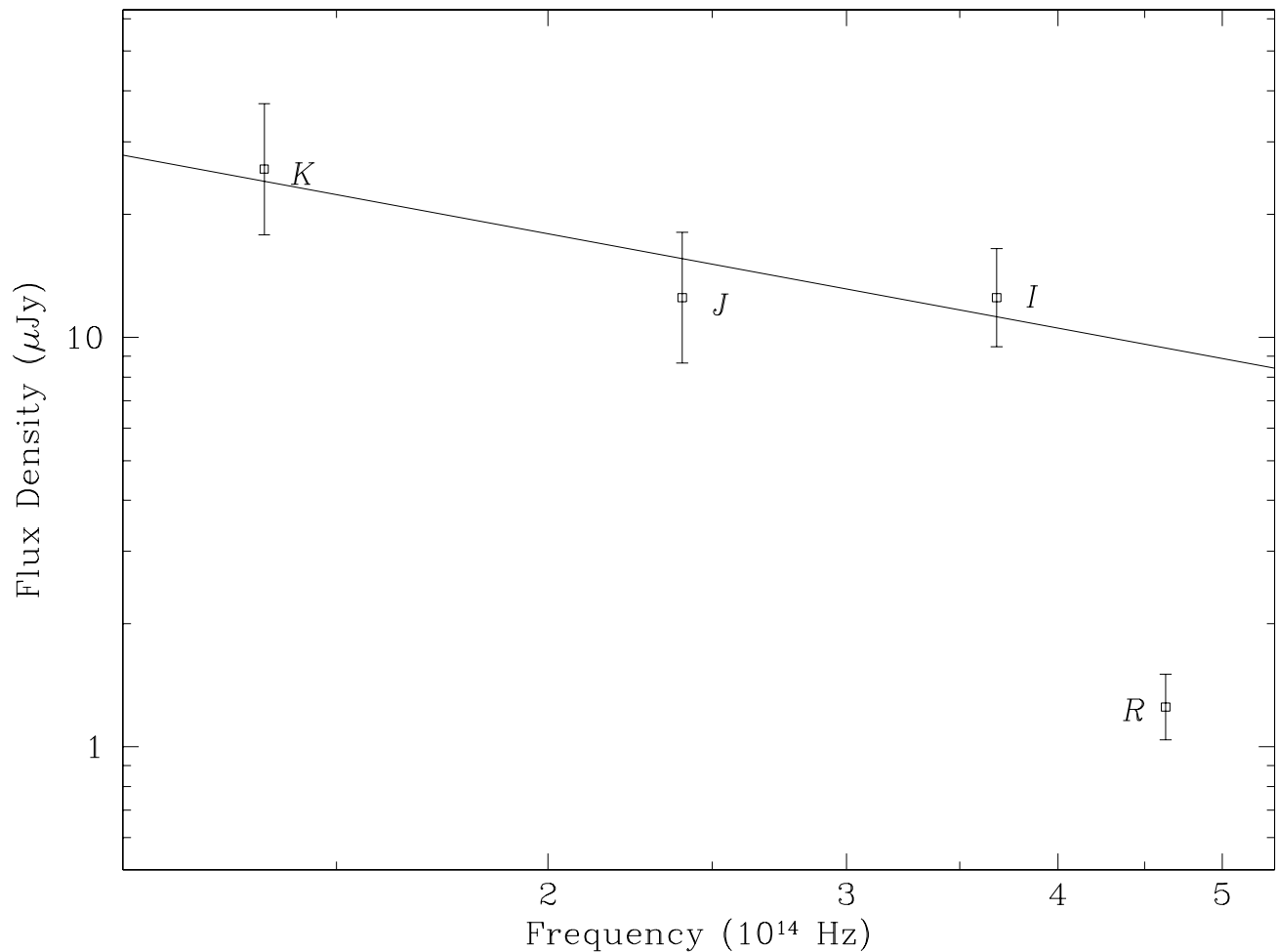,width=5.5in,angle=-90}}
\vspace{10pt}
\caption{The spectral energy distribution of the OT associated with
GRB~980329 estimated for 29.9 March 1998.  The $I$ and $R$ bands
were observed at that date; the $K$ and $J$ bands have been extrapolated
from observations at later times (see text).  The diagonal line
shows the spectral slope expected from synchrotron
emission in an expanding fireball with the observed temporal
behavior.}  
\end{figure}

In addition to the Galactic extinction adjusted
flux densities of the OT, the figure
displays a 
sloping line of the form $\nu^{-0.76}$.   
If, as the temporal index indicates, our observing frequency
is above $\nu_m$ (the frequency associated with the
minimum specific energy $\gamma_m$ imparted by the shock to the
radiating electrons),  then the expected (unextincted) spectral slope
is $ 2/3 $ the temporal index of $-1.14$, or $0.76$ \cite{wrm97}.   The
$K$, $J$ and $I$ magnitudes clearly fit the spectral slope well,
while the $R$ magnitude is about a factor of $7 \pm 2$ below
the line.     While it is hard to understand how dust could produce
a decrement of a factor of $ \sim 7$ in $R-I$ while leaving $J-I$
unaffected, there is a well-observed astrophysical phenomenon
completely consistent with the observations: the absorption
of light by the Ly$\alpha$ forest.  
At high redshift intergalactic
clouds of hydrogen cross any random line of sight sufficiently
frequently to significantly reduce the observed far UV continuum 
of a background source.   The strength of this absorption depends
on the frequency of the light and the redshift of the background
source.  At redshifts $\simlt 2$, the absorption primarily occurs
at rest wavelengths shorter than the Lyman limit, $912 \AA$;
however, at higher redshifts, the Ly$\alpha$ forest becomes
sufficiently dense that rest wavelengths up to Ly$\alpha$ ($1216 \AA$)
become significantly depressed \cite{mad95}.  For a $z \sim 5$ object, 
this depression, or Ly$\alpha$ break, occurs in the observer's
frame at $ \sim 7300 \AA$.  
Thus $z \sim 5$ objects 
will be $R$ ``dropouts". 
Indeed, Madau (1995) predicts
a suppression of $80 \%$ of the flux of an object at $z \sim 5$
shorter than $1216 \AA$,
and observations of spectroscopically confirmed $z \sim 5$ galaxies
\cite{wsb+98,ssb+98} 
agree well with this prediction (see Figure 4 in Spinrad et al. 1998),
with the observed depression perhaps slightly greater than predicted
($\sim 90 \%$).
Therefore not only does the wavelength of the break in the OT
spectrum agree with a $z \sim 5$, but also the magnitude of the
break agrees with that observed.  It should perhaps also be noted
that the ``dropout'' technique has worked successfully not only at
$z \sim 5$ , but also at ``lower" redshifts ($2 \simlt z \simlt 4$) \cite{sgpd+96,mfd+96,lkg+97},
and thus is a well-tested technique for obtaining photometric redshifts.

To quantify the range of redshifts consistent with the data, one can
convolve a relatively typical CCD response curve (the STIS CCD was used)
with the Harris R filter transmission
function (KPNO filter \#1466) to simulate
the full $R$ response of the observing telescopes.
One finds that if the Ly$\alpha$ depression
is 90\%, then the break spectrum would have to be at $\lambda = 7000 \AA$ 
({\it i.e.} 
$z = 4.75$) to
produce a reduction in observed $R$ of 3.5 -- still
significantly less than that observed.
But if the true Ly$\alpha$ decrement is closer to the $80\%$ predicted,
the break would have to occur out of the $R$ band and into
the $I$ (and thus have a $z > 5.2$) to have the observed colors. 

The reader may be concerned that there could be another cause
for this large spectral break.  Decrements nearly this
large have been observed due to other causes in very rare
objects ({\it e.g.} 
the 2800 \AA\,\, break in the iron low-ionization broad absorption
quasar FIRST J1555633.8+351758, Becker et al. \nocite{bgh+97} 1997).
However, the agreement of the break wavelength and the break magnitude
would then have to be entirely coincidental.   Rather, given the good
agreement between break wavelength and expected decrement, 
the possibility that GRB~980329 occurred
at $z\sim5$ must be taken seriously.

It is important to note that the fireball model does
not predict any spectral break of the magnitude of the one seen
here \cite{spn98}.  And although the lightcurve of GRB~970228 has 
shown some unpredicted but comparatively small variability
 (see, for example, Fruchter et al. 1998), in general the evidence
that OTs behave according to the fireball model is
quite good \cite{wrm97,gwb+98}.  

There is yet another piece of circumstantial evidence which suggests
GRB~980329 may be at $z \sim 5$.  The temporal decay
of the OT only fits a power-law
if one assumes that nearly all of the observed flux in the $R$ band
observations is from the OT and thus is not significantly contaminated
by the host.  Indeed, even a host as faint as $R = 26.5$ would
cause a significant distortion of the power-law decay.  However, only
one other GRB host is this faint 
(see Hogg and Fruchter \nocite{hf98}
1998 for a review of host magnitudes) -- that of GRB~971214.   This
host has a spectroscopically measured redshift of
$z = 3.4$ \cite{kdr+98} and has $V = 26.5$  \cite{odk+98}.  Yet
with $z = 3.4$ one would expect the $V$ magnitude to be partially suppressed by
Ly$\alpha$ absorption; thus the source is probably intrinsically
brighter.   Furthermore, 
the apparent $M_*$ (the knee of the luminosity function) 
for galaxies at $z \sim 5$ is expected to be approximately
26 mag \cite{hf98}.  
The limit on the host magnitude
is already fainter than an $M_*$ galaxy
at $z \sim 5$.

\section{Discussion}

Were GRB~980329 truly at $z \sim 5$, the implications for our understanding
of the energetics and beaming of GRBs would be profound.  
As noted by in 't Zand et al. (1998) \nocite{zaa+98} the
fluence of GRB~980329 in the 50-300 keV of $2.6 \times 10^{-5}$ ergs
s$^{-1}$ cm$^{-2}$ would place it in the top 4\% of GRBs in the
Batse 4B catalog \cite{pmp+97,mpp+98}.  With an assumed cosmology of
H$_0 = 70$ km/s and $\Omega = 0.3, \Lambda = 0$ this fluence
implies an isotropic burst would have emitted $\sim 5 \times 10^{54}$
ergs in gamma rays alone.  This is equivalent to the rest mass
of a 2 $\Msun$ object, and would therefore imply strong beaming
for any of the GRB mechanisms discussed earlier. 

However, the implications of a $z \sim 5$ burst may be nearly as great
for cosmology as for the GRB field.   GRB~980329
is only one of 9 bursts with well-identified optical transients
(see Hogg and Fruchter 1998).  This result would imply that
$\sim 20 \% $ of bursts are at $z > 3$ and $\sim 10 \%$ of
bursts are at $z \simgt 5$.  
And, if GRB~980239 is any indication,
these high-redshift bursts will fairly frequently be bright enough
to be detected in the optical by a 1-m telescope!  Indeed, the
apparent magnitude of these objects could allow unprecedented
studies of the galactic and intergalactic medium at  extremely high
redshifts through high resolution spectroscopy either from 10-m
class telescopes on the ground
or from NGST.

Although at the time of this writing the OT should have faded by
a factor of $\sim 500$ from the
flux densities shown in Figure~1, the opportunity to measure the
redshift of GRB~980329 may not be lost.  The host galaxies
of all other GRBs with confirmed OTs have been found.  If GRB~980329
also has a host galaxy, then the light from that host will suffer
the same attenuation in the intergalactic medium as the OT,
and the $R-I$ color of the host, like that of the
OT, should be larger than 2.   While quite possible apparent magnitudes of 
the host ($I \sim 27$, $R \sim 29$) are daunting, they are 
within the capabilities of the next imaging instrument planned
for HST, the Advanced Camera for Surveys.   Therefore, the 
surprising conclusion
of this paper, that GRB~980329 may have been
at $z \sim 5$, need not go untested.

\section{Acknowledgements}

I would like to thank Elena Pian, whose comment that the power-law decay
of the OT of GRB~980329 implied a very faint (and perhaps as yet undetected)
host led me to think carefully
 about this object.  Sylvio Klose, James Rhoads, Daniel
Reichart and Francisco Castander generously
shared their data and calibrated values
of the I-band magnitude before publication.  Steve Thorsett provided
valuable comments on the draft.  I would also like to thank
my colleagues at STScI, and, in particular, Mark Dickinson, Harry Ferguson,
Piero Madau and Massimo Stiavelli for many enjoyable and informative
conversations on the properties of high redshift galaxies.

\end{document}